# Computational Modeling of Neuronal Networks

XueJuan Zhang[1]    Jianfeng Feng[2]

[1]Mathematical Department, Zhejiang Normal University, Jinhua, Zhejiang, P.R. China, 321004

xuejuanzhang@gmail.com

[2]Centre for Computational Systems Biology, Fudan University, Shanghai, PR China

Department of Computer Science, Warwick University, Coventry UK

jianfeng.feng@warwick.ac.uk (corresponding author)

## 1. Introduction

Human brain contains about 10 billion neurons, each of which has about 10~10,000 nerve endings from which neurotransmitters are released in response to incoming spikes, and the released neurotransmitters then bind to receptors located in the postsynaptic neurons. However, individually, neurons are noisy and synaptic release is in general unreliable. But groups of neurons that are arranged in specialized modules can collectively perform complex information processing tasks robustly and reliably. How functionally groups of neurons perform behavioural related tasks crucial rely on a coherent organization of dynamics from membrane ionic kinetics to synaptic coupling of the network and dynamics of rhythmic oscillations that are tightly linked to behavioural state.

To capture essential features of the biological system at multiple spatial-temporal scales, it is important to construct a suitable computational model that is closely or solely based on experimental data. Depending on what one wants to understand, these models can either be very functional and biologically realistic descriptions with thousands of coupled differential equations (Hodgkin-Huxley type) or greatly simplified caricatures (integrate-and-fire type) which are useful for studying large interconnected networks.

## 2. Single Spiking Neuron

The single neuron is the fundamental building block of every nervous system. Despite varying more or less in size and shape, almost all neurons have three basic parts: a cell body (also called soma), dendrites and an axon. Dendrites plays a role of "input receiver" that collects signals from other neurons and transmit to soma, and the soma is the "central information processer", while the axon functions as an "information sender". The cell membrane of the axon and soma contain voltage-gated ion channels. These channels are selective and only permeable to one particular type of ions, and the opening and closing of a single channel depend on the membrane voltage, ligand or second messages and are intrinsically stochastic. Here a single neuron of Hodgkin-Huxley (HH) type with a finite number of channels is considered to show how various ion channels contribute to spike generations.

Neurons communicate information via synapse, a junction that allows a neuron to pass electrical or chemical signals to one or another. Classically, neurons receive synaptic inputs from other neurons via their dendrites, and encode it as very brief electric events (action potentials, also called spikes). These spikes are propagated along nerve axons to the nerve endings, where the signals are transmitted to other neurons via release of neurotransmitters at synapses. Most neurons in the central nervous system (CNS) use either the excitatory transmitter glutamate (AMPA, α-amino-3-hydroxy-5-methyl-4-isoxazolepropionic acid; NMDA, *N*-Methyl-D-aspartic acid) or the inhibitory transmitter GABA (gamma-Aminobutyric acid), but some use a relatively small number of alternative neurotransmitters, the best known of which are acetylcholine, histamine, noradrenaline, serotonin and dopamine et al.

## 2.1. The Hodgkin-Huxley (HH) model

Instead of considering the deterministic HH model (Hodgkin and Huxley 1952), here a stochastic version of the HH model is presented to account for the stochastic effect of finite number of ion channels (Neher and Sakmann 1976). Consider a certain patch of membrane with an area, on which there are $N_K = \rho_K \times Area$ potassium channels and $N_{Na} = \rho_{Na} \times Area$ sodium channels. The corresponding membrane potential *V(t)* is described by

$$C_m \frac{dV}{dt} = -[g_L(V-V_L) + g_K(V,t)\cdot(V-V_K) + g_{Na}(V,t)\cdot(V-V_{Na})] + I_{app}. \qquad (1)$$

where $C_m$ is the membrane capacity, $V_L$ is the resting leaky potential, $V_K$ and $V_{Na}$ are the reverse potentials for potassium and sodium respectively. $I_{app}$ is the external input; $g_K(V,t)$ and $g_{Na}(V,t)$ are voltage-dependent potassium and sodium conductance given by

$$g_K(V,t) = \hat{g}_K \frac{O_K(V,t)}{N_K} \;,\; g_{Na}(V,t) = \hat{g}_{Na} \frac{O_{Na}(V,t)}{N_{Na}} \;. \qquad (2)$$

In Eq. (2), $\hat{g}_K$ and $\hat{g}_{Na}$ are the potassium conductance per channel and the sodium conductance per channel, respectively; $O_K(V,t)$ and $O_{Na}(V,t)$ are the number of open potassium and sodium channels, both are voltage-dependent. Each sodium channel is composed of three identical m active gates and one h inactive gate, and each potassium channel is composed of four identical n active gates.

Denote $(m_i, h_j)$ be the state of a sodium channel with $i$ m gates and $j$ h gates open, then the state evolution of the sodium channel can be considered as a Markov process, with kinetics being given by

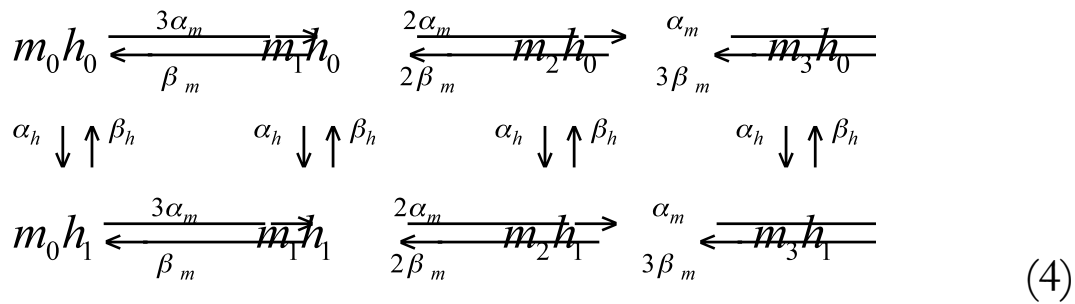

(4)

where all α,β are transition rate from one state to another, usually depending on the current membrane potential. Similarly, the kinetics of the potassium channel is characterized by

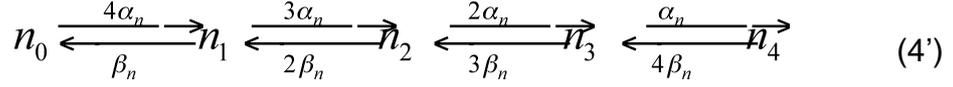

$$n_0 \underset{\beta_n}{\overset{4\alpha_n}{\rightleftarrows}} n_1 \underset{2\beta_n}{\overset{3\alpha_n}{\rightleftarrows}} n_2 \underset{3\beta_n}{\overset{2\alpha_n}{\rightleftarrows}} n_3 \underset{4\beta_n}{\overset{\alpha_n}{\rightleftarrows}} n_4 \quad (4')$$

Corresponding to the kinetics of a single sodium channel, the matrices of transition rates for m gate and h gate are

$$Q_m = (Q_m(i,j)) = \begin{bmatrix} -3\alpha_m & 3\alpha_m & 0 & 0 \\ \beta_m & -\beta_m - 2\alpha_m & 2\alpha_m & 0 \\ 0 & 2\beta_m & -2\beta_m - \alpha_m & \alpha_m \\ 0 & 0 & 3\beta_m & -3\beta_m \end{bmatrix}, \quad (5)$$

here i,j=0($m_0$),1($m_1$),2($m_2$),3($m_3$); and

$$Q_h = (Q_h(i,j)) = \begin{bmatrix} -\alpha_h & \alpha_h \\ \beta_h & -\beta_h \end{bmatrix}, \quad (6)$$

with i,j=0($h_0$),1($h_1$), respectively (see Eq. (4) and (4')). A similar transition matrix $Q_n$ can be defined for a single potassium channel.

For a given membrane patch, the open number $O_{Na}(V,t)$ of the sodium channels corresponds to the number of channels in the state $(m_3, h_1)$, and the open number $O_K(V,t)$ of the potassium channels is the number of potassium channels in the state $n_4$. It is therefore necessary to consider a system whose state at time $t_0$ can be completely characterized by a vector with 13 elements

$$(N_{m_0 h_0} = x_1, \cdots, N_{m_3 h_1} = x_8, N_{n_0} = x_9, \cdots, N_{n_4} = x_{13}) \quad (7)$$

Let $\gamma_{ij} = -(Q_m(i,i) + Q_h(j,j)), i = 0,1,2,3; j = 0,1$, it represents the escape rate of a sodium channel from a state $(m_i, h_j)$. Similarly, $\zeta_k = -(Q_n(k,k)), k = 0,1,2,3,4$ represents the escape rate of a potassium channel from a state $n_k$.

$$\lambda = \sum_{j=0}^{1} \sum_{i=0}^{3} N_{m_i h_j} \gamma_{ij} + \sum_{k=0}^{4} N_{n_k} \zeta_k \quad (8)$$

is the transition rate from the given state. So the sojourn time in the above state given by Eq. (4-5) obeys an exponential distribution, i.e.,

$$P(\tau \geq t) = e^{-\lambda(t-t_0)}, \quad t > t_0 \quad (9)$$

In simulations, the sojourn time of the system at this state can be determined by picking up a random variable $u \in U(0,1]$ (uniform distribution), and then using the relationship

$$t_r = -\ln(u)/\lambda. \quad (10)$$

Then at time $t_0 + t_r$, the state of the system has to be updated.

Next, one has to determine which state the system will occupy at time $t_0 + t_r$. From the Markov diagram of the sodium and potassium channels, one can see that there are totally 28 conformational transitions with transition rate $\eta_j$:

$$(m_0, h_0) \xrightarrow{\eta_0} (m_1, h_0),$$

$$(m_1, h_0) \xrightarrow{\eta_1} (m_0, h_0),$$

$$\ldots\ldots$$

$$(m_2, h_1) \xrightarrow{\eta_{18}} (m_3, h_1),$$

$$(m_3, h_1) \xrightarrow{\eta_{19}} (m_2, h_1).$$

There are 20 pairs in total for the sodium channel (see Eq. (4) where $\eta_j$ is given) and similarly there are 8 pairs of conformational transitions in total for the potassium channel (see Eq. (4')). The probability of escaping from state $j$ during time interval $[t_0 + t_r, t_0 + t_r + dt]$ is given by

$$\frac{\eta_j dt}{\sum_{j=0}^{27} \eta_j dt} = \frac{\eta_j}{\lambda}. \tag{11}$$

Let $P_0 = 0$, $P_k = \dfrac{\sum_{j=0}^{k} \eta_j dt}{\lambda}$, $k = 0, \cdots, 27$, numerically, the state transition is realized at time $t_0 + t_r$ by picking up a uniformly distributed random number $r \in U(0,1]$, and find $k$ such that

$$P_k < r < P_{k+1} \tag{12}$$

The determined $k$ is used to update the state of the system. For example, when $k = 1$, conformational transition $(m_1, h_0) \to (m_0, h_0)$ occurs, $N_{m_0 h_0}$ is then replaced by $N_{m_0 h_0} + 1$ and $N_{m_1 h_0}$ is replaced by $N_{m_1 h_0} - 1$.

Once a state transition occurs, the values of $N_{m_3 h_1}$ and $N_{n_4}$ are recorded, and the membrane potential $V(t)$ is updated, accordingly. This numerical method is called Gillespie algorithm (Gillespie 1976). One can instead use a Monte Carlo method by tracking the states of each individual channels. But the later one is much more time consuming if the number of channels is large.

Although the importance of channel noise on information transmission between neurons is not fully understood, it has been shown that the stochastic flickering of ion channels is crucial for spontaneous firing (Chow and White 1996) and subthreshold oscillations (Dorval and White 2005). Fig. 1 shows the spontaneous firing without an external input. It is seen that even without any external injected current, a single neuron can spontaneously fire due to the intrinsic noise aroused by stochastic opening and closing of ion channels with finite number. It is also seen that with the increase number of the channels, the firing rate of the spontaneous firing decreases. Actually, if there is sufficient large number of channels, the single neuron without any external input will become silent, and the stochastic model is identical to the classic HH model. For the detailed link between stochastic and deterministic HH models, see (Fox and Lu 1994).

## 2.2. Integrate-and-fire (IF) Model

In 1907, long before the mechanism of spike generation is known, Lapicque developed a very simple neuron model called integrate-and-fire model to describe the basic neuron operation(Brunel and Van Rossum 2007). In this model, action potential is generated when the integrated membrane potential $V(t)$ crosses a prefixed threshold $V_{th}$. The dynamics of the membrane potential is given by

$$\begin{cases} C_m dV = -(V - V_{rest})dt + dI_{syn}(t) + I_{app}dt, & V \leq V_{th} \\ V(t) = V_{reset}, V > V_{th} \end{cases} \quad (13)$$

The synaptic input is

$$I_{syn}(t) = a\sum_{i=1}^{p} E_i - b\sum_{j=1}^{q} I_j. \quad (14)$$

where $E_i = \{E_i(t), t \geq 0\}, J_j = \{I_j(t), t \geq 0\}$ are inhomogeneous Poisson processes with rates $\lambda_{E,i}$ and $\lambda_{I,i}$, respectively (Shadlen and Newsome 1994), a>0, b>0 are the magnitudes of each EPSP and IPSP and p and q are the total number of active excitatory and inhibitory synapses. In the case there are numerous presynaptic inputs and both a and b are small, one can use diffusion approximation to approximate the synaptic inputs (Tuckwell 1988; Feng 2004). In the case a=b, p=q, the LIF model is simplified as

$$C_m dV = -(V - V_{rest})dt + \mu dt + \sigma dB(t), \quad V \leq V_{th} \quad (15)$$

With

$$\begin{cases} \mu = a\lambda(t)(1-r) \\ \sigma^2 = a^2\lambda(t)(1+r) \end{cases} \quad (16)$$

where $\lambda(t) = \sum_i \lambda_{E,i}$ and $r\lambda(t) = \sum_i \lambda_{I,i}$, $r$ is the ratio between the inhibitory input and the excitatory input and B(t) is the standard Brownian motion.

The interspike interval (ISI) of efferent spikes is

$$T = \inf\{t : V(t) \geq V_{th}\}.$$

Despite its simplicity, to find an analytical expression of T has been an endeavour for many years. It was recently found out that the probability density of $T$, denoted as $p_{\lambda,r}(t)$, has the following form (here we assume that V$_{rest}$ =0)

$$p_{\lambda,r}(t) = \exp\left(\frac{-V_{th}^2 + 2\mu\gamma V_{th}}{2\gamma\sigma} + \frac{t}{2\gamma}\right) p^0(t) E_{0 \to V_{th}} \left\{\exp\left[-\frac{1}{2\gamma^2\sigma}\right] \int (v_s - V_{th} + \mu\gamma)^2 ds\right\} \quad (17)$$

The expression looks a bit complicated, in which $\gamma = 1/C_m$ and

$$p^0(t) = \frac{V_{th}}{\sqrt{2\pi\sigma t^3}}\exp\left(-\frac{V_{th}^2}{2t\sigma}\right). \tag{18}$$

Besides, there is a stochastic process, called 3-dimensional Bessel bridge $\{v_s\}_{0 \leq s \leq t}$ being included. Mathematically, it satisfies the following stochastic differential equation

$$dv_s = \left(\frac{V_{th} - v_s}{t - s} + \frac{\sigma}{v_s}\right)ds + \sqrt{\sigma}dB_s, 0 < s < t, v_0 = 0, v_t = V_{th}. \tag{19}$$

In Eq. (19), $E_{0 \to V_{th}}$ represents the expectation with respect to the stochastic process $\{v_s\}_{0 \leq s \leq t}$ with a starting point 0 and an ending point $V_{th}$.

One of the most puzzling issues in neuroscience is how the brain encodes and then decodes information. In the setting here, the input information is the input rate $\lambda(t)$, and the observed information is firing patterns of an ensemble of neurons (see Fig. 2). Having an exact expression of the distribution $p_{\lambda,r}(t)$ of the interspike interval, the best strategy to decode the input information is by applying the maximum likelihood estimate. The detailed strategy of this method was given in (Zhang et al. 2009). Here spikes from an ensemble of IF neurons are collected and based on the spike patterns, an external dynamic input can be accurately and reliably read out. Assuming that each neuron receives a common Poisson synaptic input (for simplicity, only excitatory input is taken into account), with a rate given by: $\lambda(t) = 2 + 4\left(\sin^2(2\pi t) + \sin^2(\frac{3}{2}\pi t)\right)$, for example, Fig. 2C depicts the decoded rate vs. the input frequency for a decoding time window of $T_W = 25ms$. Although $T_W$ is very short, one can see that the estimate is accurate (except that it is slightly downward biased). It is clearly shown that within a very short time window (~25 ms), the spikes generated from an array of neurons contain enough information to decode the input information.

## 2.3. Differences between IF and HH models

As two most commonly used neuronal models, the HH model and the IF model could behave quite differently in response to an external input. For example, these two models respond to correlated input in totally opposite ways. It has been summarized in (Feng, 2001; Feng and Zhang 2001) that increasing the input correlation to an IF neuron increases the output variability; while oppositely, the correlation of input decreases rather than increases the output variability of a HH neuron. This says that increasing the input correlation will increase the signal-to-noise ratio of the efferent spike train in the HH model, whereas for the IF neuron, the signal-to-noise is decreased with the increase of the correlated input. It was therefore hypothesized that the IF models may work better in an environment of asynchronous inputs, while the HH models is more suitable working in an environment of synchronous (correlated) input. The difference between these two models can be clearly seen from Fig. 3.

# 3. Network Model

Neurons are usually grouped in specialized modules in order to perform cognitive tasks, such as visual recognition, perception, working memory, decision making, spatial navigation etc. All these neuronal activities are marked by oscillatory rhythms whose signal properties are characterized by frequency, amplitude and phase. Using EEG recording, most of the rhythmic activities in the brain are classified into different frequency bands, the best known are: delta (1-4 Hz), theta (4-8 Hz), alpha (8-12Hz) beta (12-30 Hz) and gamma (30-70 Hz) frequency band. Neuronal network models have been extensively investigated to address the functional meaning and the corresponding mechanisms of brain oscillations. Although rhythms with different frequencies and amplitudes operate in different spatio-temporal scales, in many cases, slow brain oscillations, such as theta waves, coexists with fast brain oscillations such high-gamma waves, and there are cross-frequency modulation between slow and fast oscillations. This has been reported in both cortex and hippocampus (Buzsaki et al. 2003; Canolty et al. 2006; Tort et al. 2009). Among different types of cross-frequency interactions, nesting was proposed to be the most possible form of interactions.

To see how neuronal oscillations with different rhythms work together to generate complex brain functions, one can consider a spiking neuronal network which could generate coupled theta/gamma oscillations by utilizing a combination of fast and slow-type GABA receptor interneurons. Each neuron obeys an integrate-and-fire equation (13), it receives AMPA and NMDA receptor-mediated currents from excitatory (EX) cells, GABA receptor-mediated currents from fast inhibitory neurons as well as slow inhibitory neurons (see Fig. 4). The gating variable $s$ for AMPA and NMDA receptors is described by two first-order kinetics:

$$\frac{dx}{dt} = \alpha_x \sum_j \delta(t - t_j) - x/\tau_x, \quad \frac{ds}{dt} = \alpha_s x(1-s) - s/\tau_s, \qquad (20)$$

where $t_j$ is the presynaptic spike time. The parameters are set as follows: $\alpha_x = 1$ (in dimensionless), $\tau_x = 0.05\text{msec}, \alpha_s = 1.0\text{msec}^{-1}, \tau_s = 2.0\text{msec}$ for AMPA receptors; and $\alpha_x = 1$ (in dimensionless), $\tau_x = 2\text{msec}, \alpha_s = 1.0\text{msec}^{-1}, \tau_s = 80\text{msec}$ for NMDA receptors. The gating variable $s_{GABA}$ for GABA receptors obeys a simple first-order kinetics :

$$\frac{ds_{GABA}}{dt} = \alpha_I \sum_j \delta(t - t_j^-)(1 - s_{GABA}) - s_{GABA}/\tau_I, \qquad (21)$$

where $t_j^-$ indicates the time immediately before the spike at time $t_j$. Here $\tau_I = 9\text{msec}, \alpha_I = 1\text{msec}^{-1}$ for the fast GABA channels, and $\tau_I = 50\text{msec}, \alpha_I = 0.2\text{msec}^{-1}$ for the slow GABA channels. The AMPA and NMDA receptors-mediated currents are given by: $I_{AMPA} = g_{AMPA} s_{AMPA}(V - V_E)$ and $I_{NMDA} = g_{NMDA} s_{NMDA} B(V)(V - V_E)$, respectively, with $B(V) = [1 + \exp(-0.062V)/3.57]^{-1}$. The GABA receptor-mediated current is given by $I_{GABA} = g_{GABA} s_{GABA} (V - V_I)$. Here $V_E = 0\text{mV}$, $V_I = -70\text{mV}$.

Diagram of the spiking neural network which incorporates two kinetics of GABA current is depicted in Fig. 4A. The firing pattern of the network in response to a square-shaped stimulus was shown in Fig. 4B, where excitatory neurons which receive synaptic currents from slow and fast inhibitory neurons are shown to exhibit the theta-nested gamma rhythm, and the stimulus-evoked response was clearly demonstrated. By using a Morlet wavelet in the range of frequency between 1Hz and 100Hz, it was shown that the corresponding spectral component to onset of stimulus is highly concentrated in a low frequency band (around 4-8Hz which is referred as theta band, see the second trace of Fig. 4C). Besides strong powers in the theta band, there are weak powers in the gamma band (30-80Hz). Furthermore, it was seen that when the stimulus is on, both theta and gamma powers are increased, but the magnitude of theta wave in the presence of stimulus is much higher than that of gamma wave (see the bottom trace of Fig. 4C).

Some studies reported that the coupling between theta and gamma may depend on the theta phase rather than its amplitude (Buzsaki 2006; Rizzuto et al., 2003); while some revealed that the magnitudes of both theta and gamma oscillations are involved to predict the efficacy of the subsequent meory recall (Sederberg et al., 2003). It was also reported that the theta oscillation can both modulate gamma amplitude and the firing of single neuron. Recently, experimental studies on the frontal and temporal lobes, rat hippocampus during olfactory learning, revealed that the modulation of theta phase to gamma amplitude is enhanced (Canolty et al. 2006, Tort et al. 2009). Simultaneous local field potential and multi-unit neuronal activity recordings in the inferotemporal cortex of conscious sheep further revealed that the amplitude of theta (but not gamma) and coupling between theta phase and gamma amplitude are enhanced following learning (Kendrick et al. 2009). The synaptic mechanism of the learning-altered theta/gamma interaction can be explained by the computational model described above. The computational studies suggested that the strengthened modulation of theta phase to gamma amplitude requires co-ordinately increase of both NMDA receptors and GABA slow receptors (Kendrick et al., 2009). This conclusion is in consistent with another modelling study on hippocampal population activity during sequence learning and place field placement, where it was shown that the conductance of GABA slow receptor which is responsible for the generation of the theta rhythm rises and falls in a rhythmic fashion with the theta rhythm (Wallenstein and Hasselmo 1997).

There is another similar spiking neuronal network that incorporates two kinetically different GABA receptor-mediated currents to study the role of the theta rhythm in decision making. To mimic decision making between two choices A and B, a computational model is generally composed of a population of excitatory neurons and a population of inhibitory neurons (Fig. 5). Intuitively, the network has two attractors (attractor A represents decision A and attractor B represents decision B), both are specified by two pools of excitatory neurons. Smerieri et al.

(Smerieri et al., 2010) compared three versions of the inhibitory population: i) only one pool of fast GABA receptor interneurons with a GABA time constant of $\tau_{GABA,fast}=10ms$ (one pool fast model); ii). two pools of interneurons, the diagram of such a two-pool network is depicted in Fig. 5A, where a fraction $S$ (for example $S=0.25$) of the inhibitory neurons has $\tau_{GABA,slow}=100ms$; iii). only one pool of slow GABA receptor interneurons. It was found that although the average time-integrated change of conductivity of the three models are the same, the reaction time of the two-pools network is considerably reduced in decision making, compared with the previous two networks with only one pool of interneurons (see Fig. 5 B). To see the difference between the firing behaviors of one pool and two-polls networks more clearly, comparison of the reaction times of networks with only one fast pool of inhibitory neurons and with two pools of fast and slow inhibitory neurons is shown in Fig. 5C. It is clearly seen the advantage of the two pool model in the reaction time of the decision making. Having a further comparison of the average firing rates before the decision cues are applied, the two pool model shows a higher firing rate, on average, for the neurons that win in the competition than for the neurons in the slow and fast one pool model. Moreover, the average membrane potentials of the two pool model are higher than the other two models.

## 4. Summary

Two classes of single neuron model are most widely used in the literature: the abstract model of IF type and biophysical model of HH type. Choosing which type model in research critically depends on the scientific question one wants to address and the experimental data in hand. Ideally one should have all voltage clamped data of each channel of the cells in the neuronal network, it is natural to use the HH type of the model by carefully match the model output with the experimental recordings. Unfortunately, it is very rare to have all the data available and one has to resort to using IF type model.

- Cross-References
- →00369  Patch clamp recording of single channel activity - acquisition and analysis
- →00358 Potassium channels (Kv, Kir KCa and K(2P) channels
- → 00356 Bioelectricity, ionic basis of membrane potentials, propagation of voltage sigMneadlis

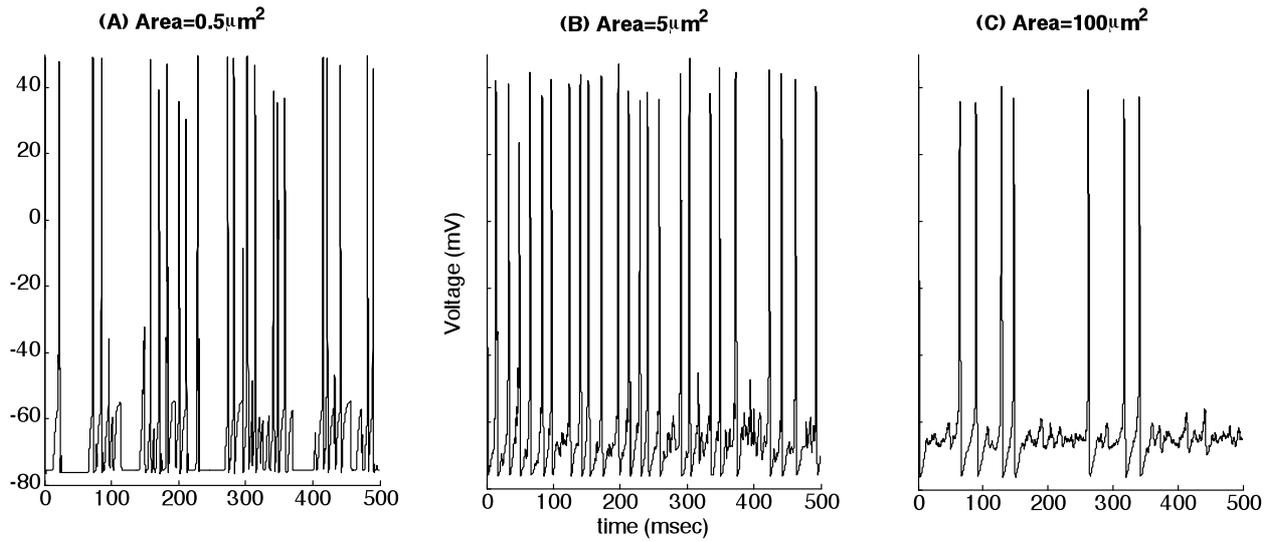

Figure 1. The time evolutions of the membrane voltage of the stochastic HH model in the absence of an external input current for different patch size: (A) is for $Area = 0.5 \mu m^2$, (B) is for $Area = 5 \mu m^2$ and (C) is for $Area = 100 \mu m^2$. Other parameters for the model are from (Chow and White 1996).

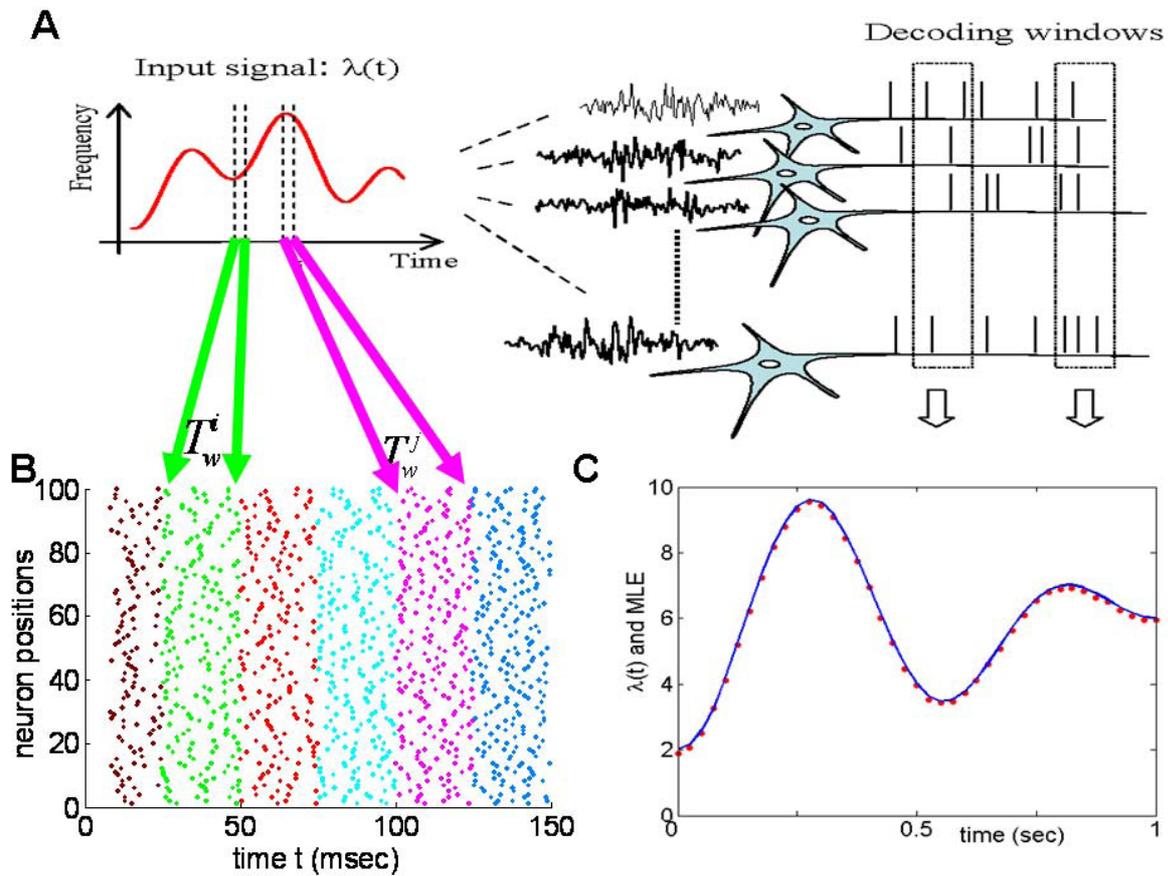

Figure 2. **MLE in a network without interactions.** (A). Schematic plot of reading dynamic inputs from an ensemble of neurons. For a fixed time window (indicated by $T_w^i, T_w^j$), the spikes are collected and the input is decoded by means of maximum likelihood estimator. (B). Raster plot of spikes in five different decoding windows. The information is read out from the spikes in each window. (C). An example of reading out the dynamic input rates from an ensemble of neurons is shown. The original signal $\lambda$ is plotted in the continuous line, while dots are estimated values of $\lambda$. Adopted from (Zhang et al. 2009).

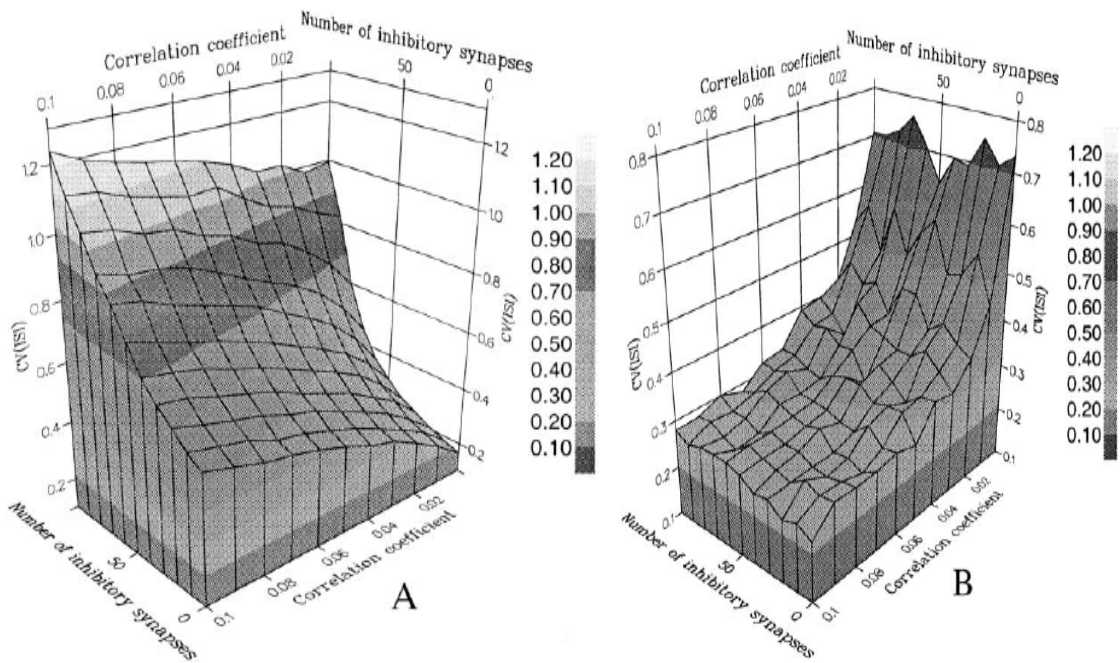

Figure 3. The coefficient of variability of efferent spike trains vs. the input correlation and the number of inhibitory synapses. (A) is the IF model, (B) is for the HH model. Adopted from (Feng and Zhang 2001).

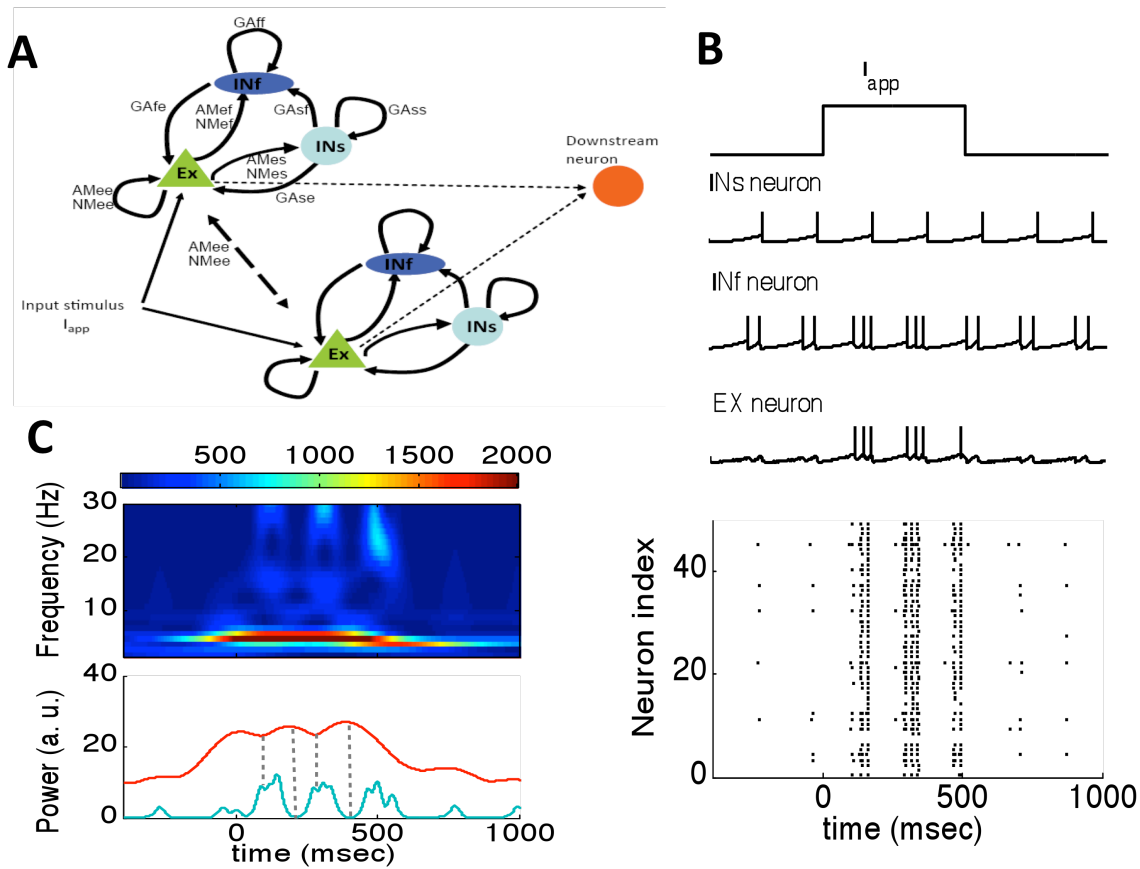

Figure 4. Neural network model showing theta/gamma dual oscillations and the response of such a network to an external input. (A) Schematic showing three populations of neurons which are connected recurrently by different synaptic receptors are the basic functional units in the neural network model. Here EX, INf and INs represents excitatory, fast inhibitory and slow inhibitory neurons. The connections, for example NMes represents the connection from Ex to INs neurons mediated by NMDA receptor. (B) Responses to a step input stimulus of a randomly chosen neuron from three populations of neuron. (C) Top: The power contribution of different frequencies across the theta/gamma range. Bottom: comparison of the mean theta and gamma powers. Here the time-frequency analysis is applied to the averaged membrane potentials of all EX neurons.

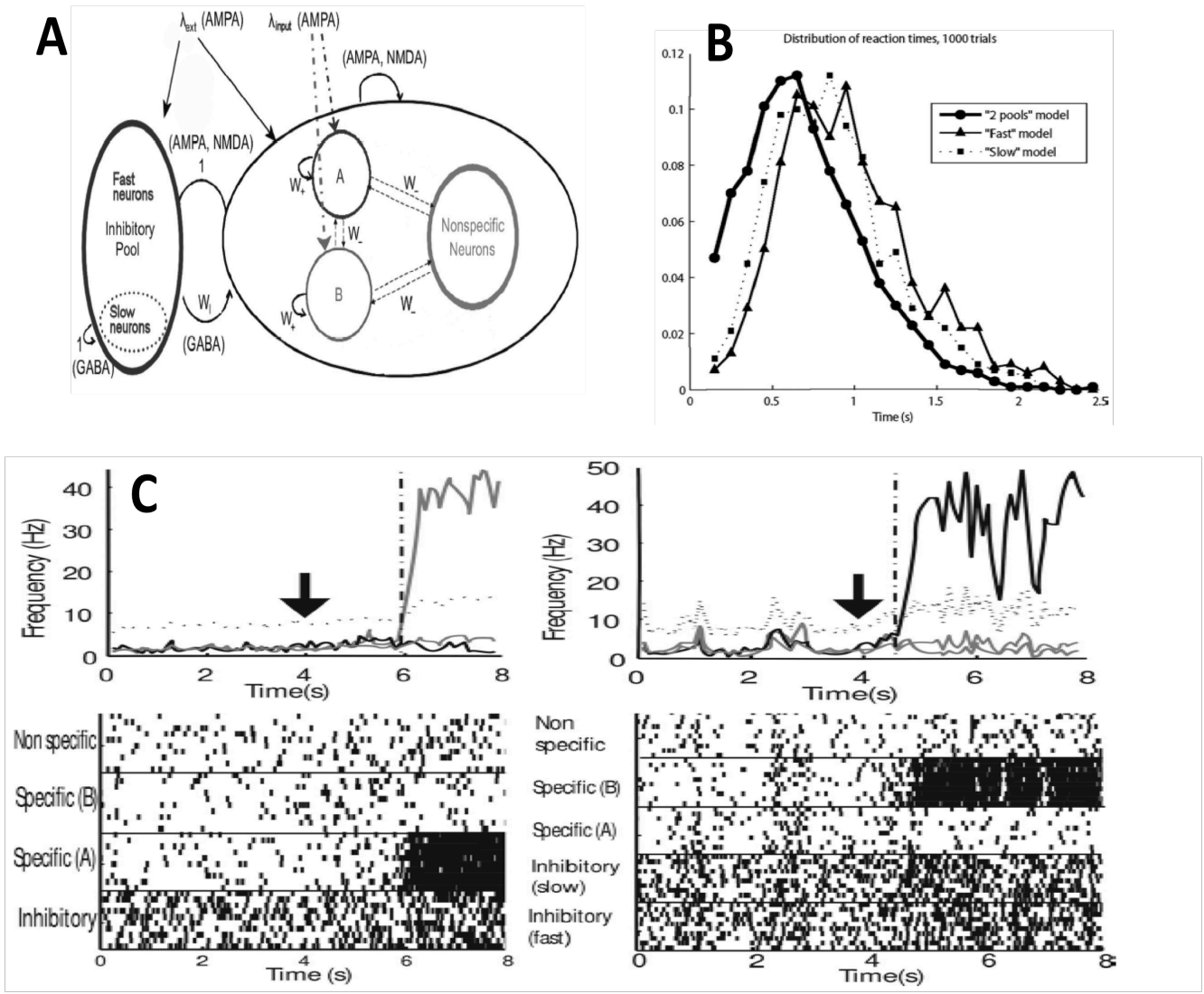

Figure 5. (A) Diagram of the two-pool network model. (B) Distribution of the reaction times for the three different models. Circles, two-pool model; triangles, fast model; squares, slow model. (C) Comparison for the firing behaviors of the one pool fast model (left) and the two pool model (right). Top, firing rates for all the pools for sample trials of the one pool fast model. Arrows mark the time of input cues applied, and the vertical dashed line marks the moment where the decision was taken. Thick gray curve and thick black curve are the rates for specific pool A and specific pool B, respectively. Bottom, Rastergrams of the spiking activity for 10 randomly chosen neurons from each population. Adopted from (Smerieri et al. 2010).